\documentclass[conference]{IEEEtran}

\usepackage[cmex10]{amsmath}
\usepackage{siunitx}
\usepackage{cite}
\usepackage{psfrag}
\usepackage[utf8]{inputenc}
\usepackage[T1]{fontenc}
\usepackage{amsmath,amsfonts,amsbsy,amssymb}
\usepackage{mathabx}
\usepackage{mathrsfs}
\usepackage[nolist]{acronym}
\usepackage{tabularx}
\usepackage{amssymb}
\usepackage{amsmath}
\usepackage{graphicx}
\usepackage{cite}
\usepackage{multirow}
\usepackage{wasysym}
\usepackage{multirow}
\usepackage{float}
\usepackage{xcolor}
\usepackage{subcaption}
\usepackage{algorithm}
\usepackage{algorithmic}
\usepackage{xcolor}
\usepackage{suffix}
\usepackage{url}
\usepackage[font=footnotesize]{caption}
\usepackage{epsfig}

\begin{document}

\title{Three-layer Approach to Detect Anomalies in Industrial Environments based  on Machine Learning}

\author{
Daniel Gutierrez-Rojas$^{1}$, Mehar Ullah$^{1}$, Ioannis T.  Christou$^{2}$, Gustavo Almeida$^{3}$, Pedro Nardelli$^{1}$, Dick Carrillo$^{1}$,\\ Jean M. Sant'Ana$^{4}$, Hirley Alves$^4$, Merim Dzaferagic$^{5}$, Alessandro Chiumento$^{5}$, Charalampos Kalalas$^{6}$\\
$^{1}$LUT University, Finland \\
$^{2}$Athens Information Technology, Greece\\
$^{3}$Federal University of Minas Gerais, Brazil\\
$^{4}$6G Flagship, University of Oulu, Finland\\
$^{5}$Trinity College Dublin, Ireland\\
$^{6}$Centre Tecnològic de Telecomunicacions de Catalunya (CTTC/CERCA), Spain
}

\maketitle

\begin{abstract}
This paper introduces a general approach to design a tailored solution to detect rare events in different industrial applications based on Internet of Things (IoT) networks and machine learning algorithms.
We propose a general framework based on three layers (physical, data and decision) that defines the possible designing options so that the rare events/anomalies can be detected ultra-reliably. 
This general framework is then applied in a well-known benchmark scenario, namely Tennessee Eastman Process.
We then analyze this benchmark under three threads related to data processes: acquisition, fusion and analytics.
Our numerical results indicate that: (i) event-driven data acquisition can significantly decrease the number of samples while filtering measurement noise, (ii) mutual information data fusion method can significantly decrease the variable spaces and (iii) quantitative association rule mining method for data analytics is effective for the rare event detection, identification and diagnosis.
These results indicates the benefits of an integrated solution that jointly considers the different levels of data processing following the proposed general three layer framework, including details of the communication network and computing platform to be employed.
\end{abstract}


\section{Introduction}
Detection and prediction of anomalies in industrial environments are important for both economic and security reasons.
However, these tasks are far from trivial since anomalies are usually \textit{rare events} within datasets so that most existing algorithms fail to identify them with (ultra-)reliability, either favoring false-alarms or misdetections \cite{he2009learning,krawczyk2016learning}. 
Even worse, in the special cases where effective solutions can be found, generalization is not straightforward.
The challenge situation becomes: general approaches usually lead to false alarms and misdetections while effective solutions are very particular and cannot offer direct guidelines to other cases.

In this paper we deal with this problem by proposing a general frame to model a wide range of cases based on the advances in Industrial Internet of Things (IIoT) networks and Machine Learning (ML) algorithms \cite{nardelli2019framework}. 
In particular, we approach the problem by using a theory that considers three autonomous (but strongly dependent) layers of cyber-physical systems (CPS), namely physical, data and regulatory.
By doing so, we are capable of analyzing in a more general way the steps of data acquisition, transmission, fusion and analytic that will allow an effective anomaly detection.
Before going into details, we provide next a brief review of the state-of-the-art in industrial CPS and rare event detection.

Industrial CPS have been studied for many years, including already several deployed solutions \cite{leitao2016smart,oks2017application}.
Most of the current research focuses on how to incorporate data flows so the industrial physical processes can run in a more efficient way, particularly focusing on multi-agent systems and the concept of digital twins.
The most promising solutions would involve real-time monitoring and control \cite{yin2019real}, industrial edge computing \cite{dai2019industrial} and software-defined wireless communications \cite{hellstrom2019software}.

When dealing with CPS \cite{greer1900cyber}, three basic steps (in addition to transmission) are taken in relation to data: acquisition, fusion and analytics.
In the data acquisition phase, sensors map physical processes into data, which can be sampled based on periodic measurements (e.g., sample every second), or event-driven ones (e.g., sample every threshold crossing), or a hybrid between both (e.g., \cite{mazo2011decentralized,miskowicz2015event,ge2019distributed}).
In the fusion phase, acquired data shall be structured, disseminated and stored \cite{hall1997introduction}.
In this phase, heterogeneous data streams might be compressed/aggregated via ML algorithms.
This phase includes possible issues related to communications and also communication network technologies including low-power networks, (beyond) 5G and IoT platforms \cite{alam2017data,cheng2018industrial}.
The analytics phase is also related to the ML algorithms that are now designed to detect or predict particular patterns or events \cite{lee2011integrating,yin2012comparison}; in particular the algorithms based on associative rules have been studied to identify anomalies and rare events with high performance \cite{adamo2012data,christou2019avoiding}.
Data fusion and analytics are also related to the computing paradigm to be employed, particularly cloud or edge \cite{dai2019industrial}.

\begin{figure*}[t]
\centering
  \includegraphics[width=2\columnwidth]{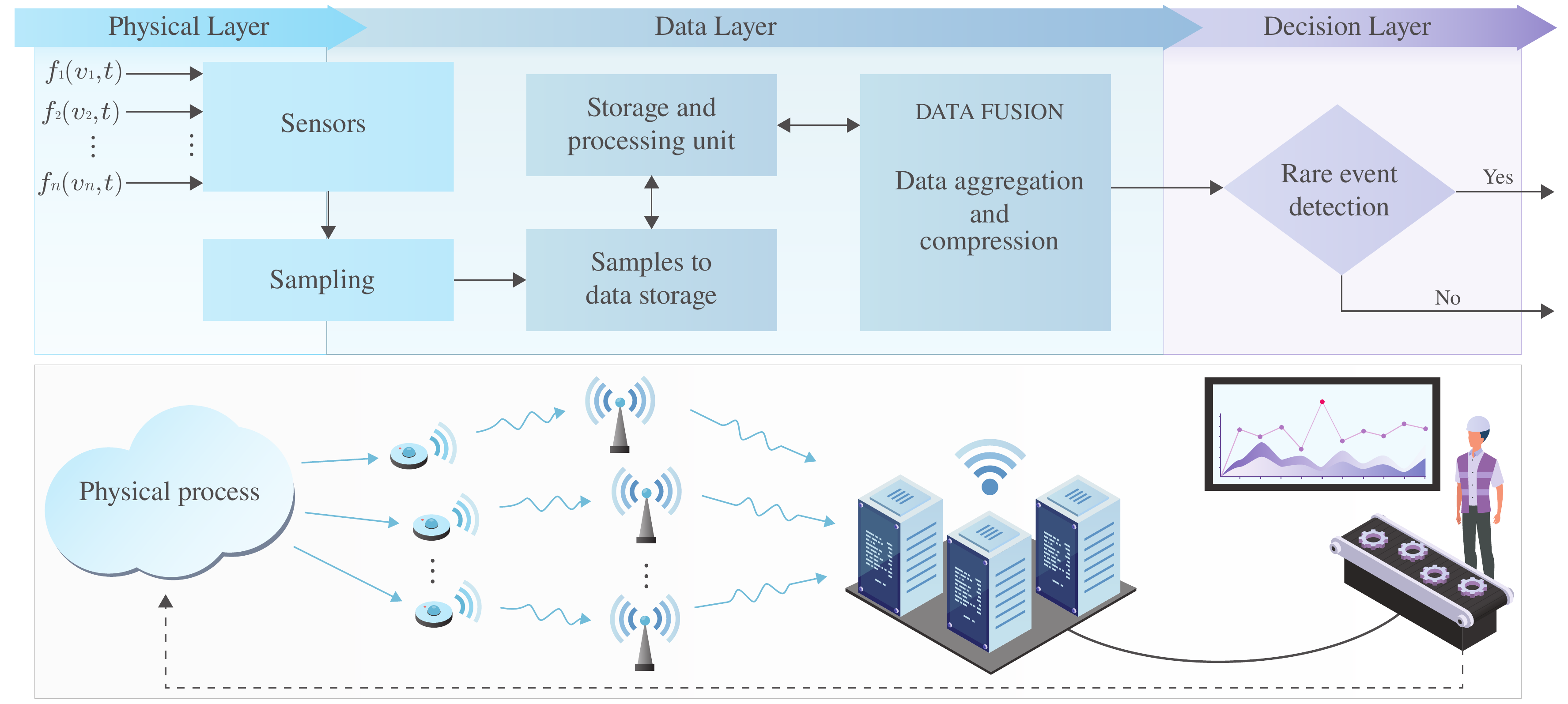}
  \caption{Three layer framework for rare event detection.}
    \label{fig:layer}
    \vspace{1ex}
\end{figure*}
This paper focus on the combination of these three basic steps following a generalized framework that defines the boundary conditions that the proposed solution for anomaly detection shall be designed.

Our specific contributions are:
\begin{itemize}
    \item Propose a general framework based on the 3-layer model of CPS, as presented in Fig. \ref{fig:layer}, from where a set of general questions related to the data acquisition, fusion and analytic steps are defined (Sec.\ref{sec:framework}).
    \item Employ the proposed framework in a well-known benchmark scenario from process engineering called Tennessee Eastman Process (TEP) \cite{russell2000fault,yin2012comparison,zhang2020incipient} (Sec. \ref{sec:tep}).
    \item Evaluate numerically the benefits of: event-driven data acquisition, mutual information  data fusion and quantitative association rule data analytics (Sec. \ref{sec:res}).
    \item Discuss different aspects about communication networks and computing paradigm usually considering perfect for this kind of studies (Sec. \ref{sec:fut}).
\end{itemize}

\section{Proposed framework}
\label{sec:framework}
There are several methods for detecting anomalies and faults in industrial settings based on data. 
However, with the steep growth of information and communication technologies, more sensors with lower time granularity are becoming common-place, generating the so-called big data. 
This, however, brings a generalized misunderstanding: ``the bigger data, the better".
In particular, in situations where the target is to reliably detect rare events in the dataset, the situation becomes more critical since the goal is to identify outliers with minimum chances of false alarm and misdetection.

\begin{table*}[t]
\centering
\caption{Related questions that can be answered by the proposed framework}
\begin{tabular}{l|l|l}

\textit{\textbf{Q\#}} & \textbf{Topic}          & \textbf{Related Question}                                                                                              \\ \hline \hline
\textit{Q0}   & { Anomaly} & What is the problem? Is the Rare Event known or unknown?                                                               \\ \hline
\textit{Q1}   & { Sensors}           & What kinds of sensors will be used? How many of each can be used and where they can be located?                        \\ \hline
\textit{Q2}   & { Sampling}          & Which type of sampling will be used? Periodic, event driven or mixed (hybrid)?                                  \\ \hline
\textit{Q3}   & { Communication}     & Which type of communication system (access and network technologies) will be used? \\ \hline
\textit{Q4}   & { Data storage}      & Where the data from sensors are stored and processed?                                                                  \\ \hline
\textit{Q5}   & { Data fusion}       & How the data should be clustered/aggregated/structured/suppressed?                                                                \\ \hline
\textit{Q6}   & { Event detection}   & How to make the ultra-reliable rare event detection based on ML algorithms?                                                   \\ 
\end{tabular}
\label{table:questions}
\vspace{-3ex}
\end{table*}

Under this new condition, the usual fragmentation between data acquisition, fusion and analytic steps needs to be reviewed.
For instance, the analytics algorithm requires structured good quality data (not necessarily ``big data") to more reliably detect anomalies.
The structured (time stamped) data, in their turn, is attained in the fusion step by disseminating, aggregating and/or eliminating data based on the needs of the rare-event detection algorithm.
But, before data is fused, it needs to be acquired by sensors that map different (usually well-defined) physical processes; sensors may be located in different places (spatial domain) and sampled (temporal domain) in potentially different ways (e.g., periodically or not). 

Our main motivation here is then to build a tailored integrated solution based on these three steps looking, for example, data compression/reduction can offer to the analytics algorithm better quality data to identify rare events.
Notwithstanding, the solution to be designed for particular shall be derived from a general framework.
In this case, the proposed framework consists of modeling CPS using three interrelated layers, namely physical, data, and decision.
This approach was proposed in \cite{kuhnlenz2016dynamics,nardelli2018smart} to assess the dynamics of physical systems that are regulated based on a decision-making processes that depend on  data processing.
The previous contributions were mainly focused on theoretical toy-models, though.
Here, we will extend this approach to focus on realistic industrial settings.
Fig. \ref{fig:layer} depicts the proposed 3-layer model, together with the underlying communication network topology.

The proposed general framework is based on key questions that must be answered before the specific solution for detecting rare-events is designed.
The questions are present in Table \ref{table:questions}.
The idea is to define the viable designing options that we need to consider to have an effective solution for a particular industrial processes, as well as practical limitations imposed by industry itself (e.g., preference for private networks, or already deployed wired communication system).

The questions are structured in steps that follow the three-layer model.
The first step is to identify the rare event(s) under consideration, also considering whether the problem is known beforehand.
To have a quantitative evaluation of the related physical processes related to the event, sensors are needed to map the physical to the data layer.
Here the question is what kind of sensors can be used? 
How many should be used and where they should be located? 
After the locations are confirmed, the next phase is the sampling strategy from those sensors: it may be periodic, event-driven (non-periodic) or a mix between them.
We then need to determine the time granularity and/or the event that trigger a sampling.
The next phase is to define the communication system to be used.
Particularly, the type of access technology (wireless or wired), the network (internet or private network) and storage (local database, cloud, private cloud).
Once the data is stored, data should be aggregated, as other variables are stored with the same timestamp. 
Depending on the information of variables monitored, some of them could be suppressed. 
The question here is how data should be clustered/aggregated/structured/suppressed (fused)?
The information gathered after the data fusion will be used to detect the event, when adding variables that monitor physical condition of the grid. Here the question arises is how to make the ultra-reliable rare event detection for our problem?
In the following section, we will briefly present the Tennessee Eastman benchmark process and then apply the proposed approach on it.

\section{Tennessee Eastman Process}
\label{sec:tep}

\begin{figure}[t]
\centering
  \includegraphics[width=\columnwidth]{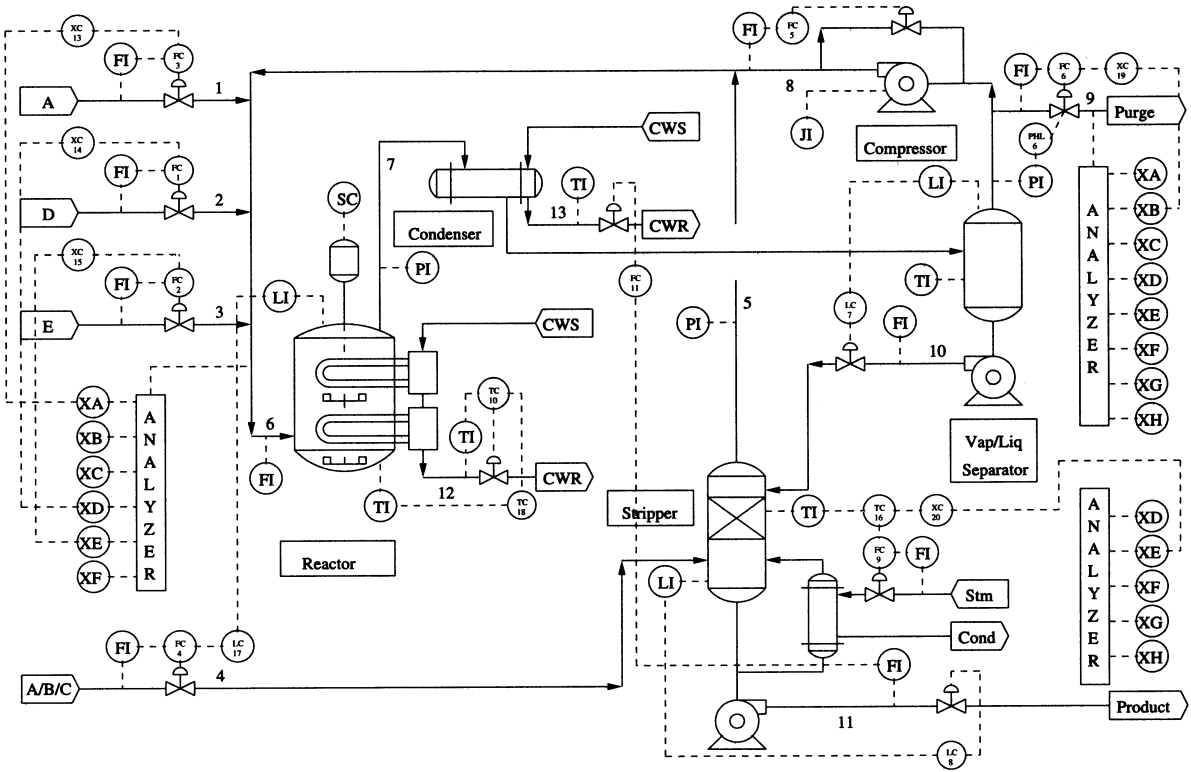}
  \caption{Process flow diagram of the Tennessee Eastman problem \cite{chiang2001}.}
    \label{fig:teppfd}
    \vspace{-1ex}
\end{figure}

The dataset material consists of several faulty cases of an industrial plant, as produced by the Tennessee Eastman problem \cite{downs1993tep}. 
The process has five major equipments, namely a condenser, a vapor-liquid separator, a reactor, a product stripper, and a recycle compressor (Fig. \ref{fig:teppfd}).
Its objective is to obtain the products G and H from the reactants A, C, D and E. This is reached by a set of four chemical reactions, in which components B and F are, respectively, an inert and a byproduct. 
More details can be found in \cite{YIN20121567,chiang2001}. %
This benchmark is suitable to evaluate process monitoring schemes and control strategies based on data driven analysis. 
Besides the normal operation, 21 abrupt or incipient faulty conditions caused by common disturbances in practice are simulated \cite{russell2000fault}. 
There are 52 monitoring variables or features, being 11 manipulated variables and 41 measured variables.
Once a faulty condition occurs, all are generally affected with changes in their respective values.

%
The Tennessee dataset was generated in a process simulator that has been widely used by the process monitoring community\footnote{\url{https://github.com/camaramm/tennessee-eastman-profBraatz}}.
It is composed by 22 subsets named \textit{dXX\_te.dat}, where \textit{XX = $0,1,2, \cdots, 21$}.
The file \textit{d00\_te.dat} refers to the normal operating condition.
Each of the other ones regards to a particular fault, that is, a different shift from this reference condition.
The subsets consist of 960 observations of the 52 variables, which are sampled every 3min with a Gaussian noise.
The faults are introduced after 8 simulation hours.
Table \ref{table:questions_TEP} presents the proposed framework applied in TEP, which provides the boundary conditions of the anomaly detection design.

\begin{table}[t]
\centering
\caption{Proposed framework applied in TEP}
\begin{tabular}{l|l}

\textit{\textbf{Q\#}} &  \textbf{Answer}                                                                                              \\ \hline \hline
\textit{Q0}   & Lack of accuracy to detect anomalies (21) from measurements \\ \hline
\textit{Q1}   & Each of the 52 variables are sensors, no other can be added                        \\ \hline
\textit{Q2}   & Periodic sampling; all 52 variables are sync (3 min.) \\ \hline
\textit{Q3}   & It can be considered as in \cite{liu2019wireless} \\ \hline
\textit{Q4}   & Not constrained; freedom to test as in \cite{dai2019industrial}
                                                                 \\ \hline
\textit{Q5}   & Open question; focus of research in the field                                                               \\ \hline
\textit{Q6}   & Open question; focus of research in the field \vspace{-2ex}                                                    \\ 
\end{tabular}
\label{table:questions_TEP}
\vspace{-2ex}
\end{table}

\section{Numerical results}
\label{sec:res}
\subsection{Event-driven data acquisition}

\begin{figure}[t]
\centering
\includegraphics[width=1\columnwidth]{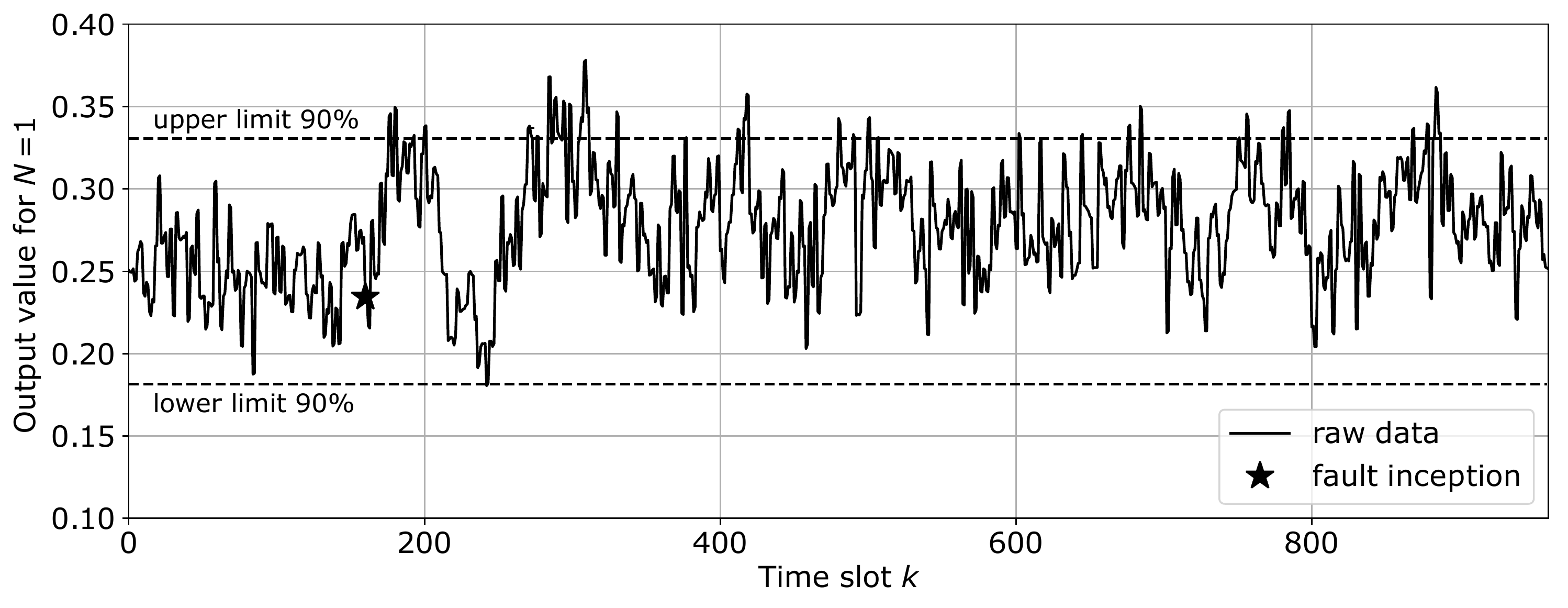}
\caption{Signal from variable 1 at data acquisition point (before transmission) for fault number 2 of Tennessee Eastman dataset.}
\label{fig:original}
\vspace{1ex}
\includegraphics[width=1\columnwidth]{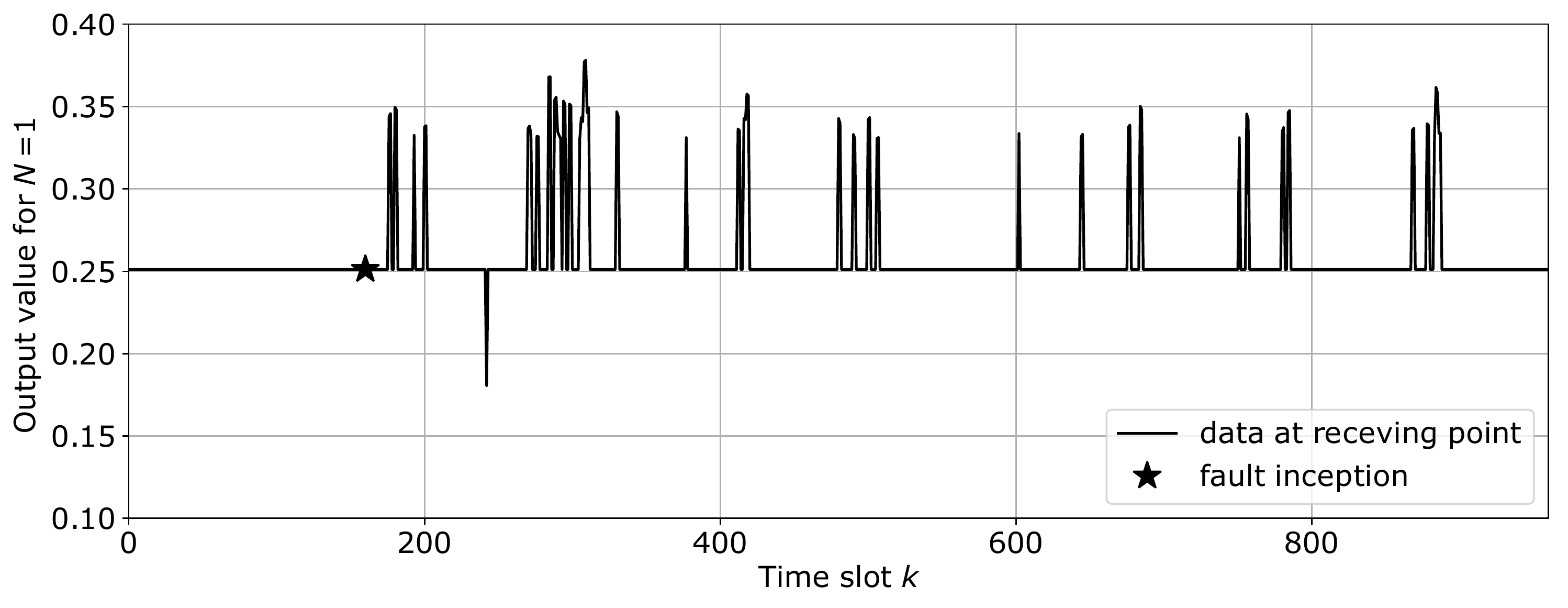}
\caption{Signal from variable 1 at data fusion point (after transmission) for fault number 2 of Tennessee Eastman dataset.}
\label{fig:event_driven}
\end{figure}

The main idea of an event-driven approach for this application is to perform data compression in order to transmit the meaning of information from the data acquisition point to the data fusion point. This approach can be described using the following steps: (1) input data from all 52 sensors ($N$); (2) variable average estimation and margin selection (90\% of lowest/highest values) from normal operation;
(3) at every time slot ($k$) for each variable, if the values are out of the margins the sample is transmitted, otherwise, if nothing is received at the data fusion point the variable will maintain the average value estimated from the previous step;
(4) compression rate calculation for each variable.
The limit values for margin selection for each variable were chosen arbitrarily. 
An example of this approach is seen in Fig. \ref{fig:original} where the signal obtained from the sensors is shown with its limits. The samples transmitted are only the ones that are out of the upper and lower limits as seen in Fig. \ref{fig:event_driven}.  

This setup allows to transmit less data via any communications system. In the example mentioned above the compression rate is 92.60\%, this means only about 7.4\% of the samples are transmitted. 
The pre-processed time series based on the proposed event-driven method will serve as inputs to the data fusion and analytics, where anomalies should be detected, identified and diagnosed.

\subsection{Data fusion based on Mutual Information}
In order to further reduce the amount of processing required for fault detection, on top of the time-series compression as proposed by the event-driven approach, the proposed framework determines how the process variables are related to each other to discover and exploit their dependencies.
The interdependencies between the process variables are determined automatically from the sampled measurements in the Tennessee datasets. 
Specifically, Mutual Information (MI) entropy reduction technique is used to infer how variables are correlated.
The MI quantifies the amount of information that each variable contains about the other ones \cite{8369004}.

\begin{figure}[t]
  \includegraphics[width=1.1\columnwidth]{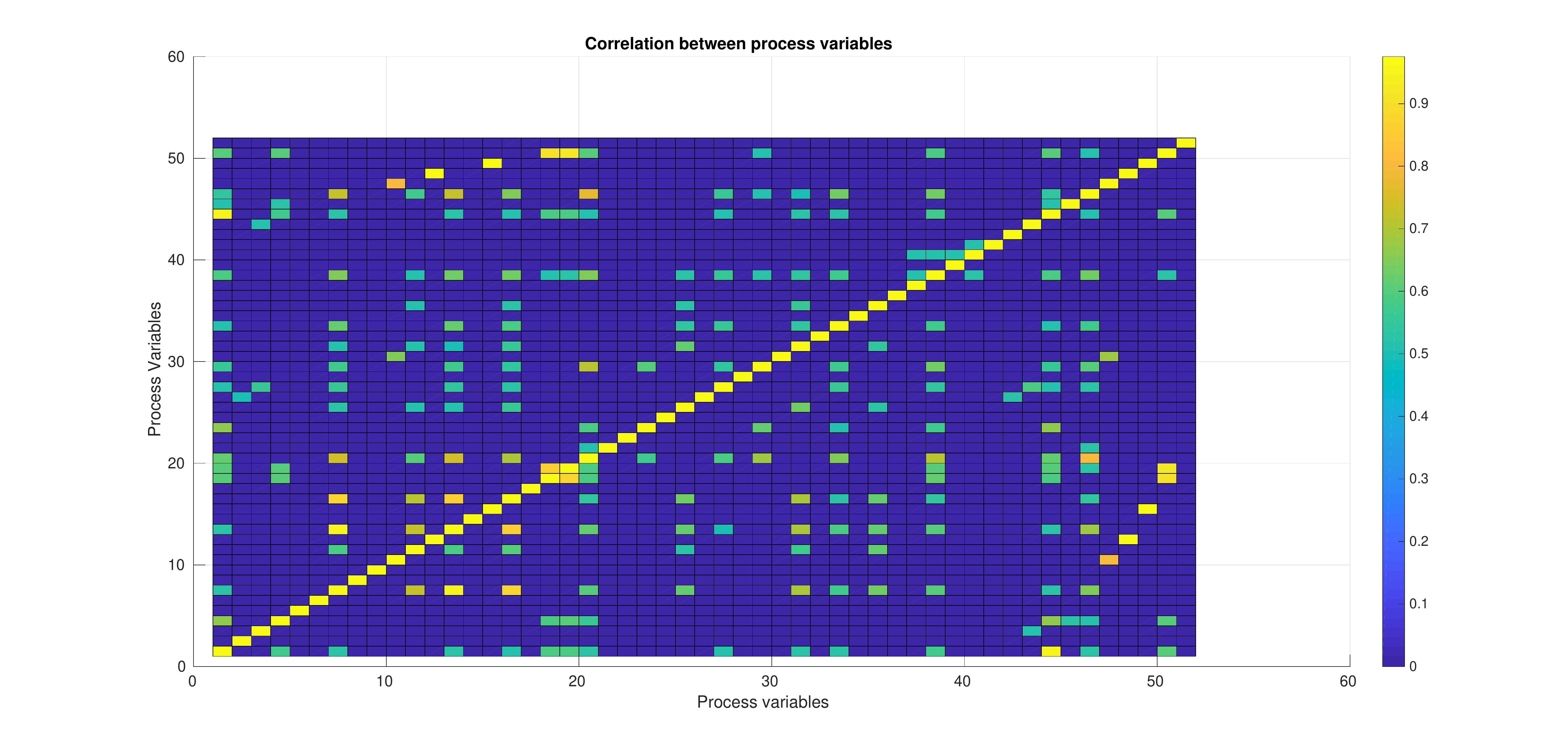}
  \caption{Correlation between Process Variables}
    \label{fig:distanceVar}
\end{figure}

The tools used to determine correlations among the variables used in this work represents the distances between variables in terms of their statistical closeness, then it quantifies the correlation by providing links between the variables. Finally, it assigns directionality to the links \cite{mider}.
Fig. \ref{fig:distanceVar} shows that aside form the auto-correlation, strong cross-correlation is present between a high number of variables; in fact, a high correlation above 80\% is present in 23\% of the variables (12 out of 52) while a modest correlation above 50\% is present in 65\% of the variables (34 out of 52).

As an example of statistical closeness, Fig. \ref{fig:distanceVar} presents the correlation for the 52 process variables in the Tennessee datasets.   
This result showcases that not every single variable needs to be observed at every given moment but, depending on the fault under investigation, it is possible to observe variables carrying a high amount of MI with other variables involved in the process. This reduces both data and computational load considerably.

\subsection{Quantitative association rule method for anomaly detection}
Quantitative association rule mining is a natural extension of classical qualitative association rule mining where the difficult task is the extraction of frequent itemsets from a dataset containing transactions. The extracted rules are statistical rules of the form $Item_1 AND Item_2 ... AND Item_k \implies Item_n$, that hold with certain support and confidence values (or other metrics) that are above user-defined thresholds. Association rule mining is one of the most heavily researched areas in data mining to this date. In our context, items correspond to the features in our dataset, and transactions are in one-to-one correspondence with the rows of the dataset; the consequent item ($Item_n$) in particular is constrained to always be the target variable in our dataset. The problem then becomes one of quantifying each item in each (qualitative) rule extracted from the dataset by constraining its value to lie in a specified interval, so that the target variable assumes a particular value. This is achieved by a modified parallel Breadth-First Search (BFS) algorithm, called QARMA, which guarantees that all (and none other) non-dominated quantitative association rules that hold in the dataset will be found (see \cite{christou2018arxiv}).

In the case of the Tennessee dataset in particular, the dataset is fully dense in the sense that every row contains values for every dataset feature, which makes the application of the qualitative association rule mining part useless. Instead, we construct all itemsets of size less than 4, and quantify each one of them separately, and in parallel, making sure that all itemsets of size $s$ are fully processed before starting to process itemsets of size $s+1$. To take into account the time-dependent nature of the data, whereby the values of any feature in the dataset are to some degree dependent on the values at the immediate previous times, we expanded the data to include for each feature, the difference between the feature's value and the feature's value in the previous 2 time-steps, resulting in a dataset having 156 different fully dense features. The QARMA algorithm took several days to run on this expanded Tennessee dataset, producing a total of $63008$ non-dominated rules, that could predict all different modes of operation except mode 0 (normal operation), and hard-to-detect faults 3, 9 and 15 (none of the 63008 rules imply faults 3, 9 or 15). The distribution of the rules among the faults is also highly skewed: faults 6 and 18 are implied by 23624 and 23884 rules respectively, whereas faults 5, 10 and 21 are implied by 9, 3, and 8 rules respectively. During testing, an instance for which more than 10 rules fire, is predicted to belong to the fault that is specified by the majority of rules firing on that instance. Using this majority vote rule ensemble, we obtained an overall rule-accuracy that exceeds 62\% on the test set. This accuracy is significantly better than the one reported by decision trees (J48), or artificial neural networks (MultiLayerPerceptron) as implemented in the WEKA ML/DM software suite, all of which reported accuracy less than 50\% on the test set. 

We also implemented another modified BFS algorithm to search to find a minimum cardinality set of variables that contain all the variables necessary for an appropriate subset of the discovered rules to cover 85\% of all instances that the entire set of rules cover; we say that "rule $r$ covers instance $i$" if and only if in the instance $i$ the values of the features that form the antecedents in rule $r$ are within the intervals specified for them by the rule, and the instance indeed belongs to the operation mode (fault number) that is predicted by the rule. Interestingly, only 14 of the 156 variables are enough to "explain" 85\% of the entire dataset covered by the rules found; this result implies that possibly a much smaller set of variables need be monitored in order to derive safe conclusions about the state of the process. The total rules found covered more than 70\% of the training set. 

We consider our first results as encouraging in that rules using only up to 2 features at a time are able to form an ensemble of rules that outperformed other well-known ML algorithms in test-set performance. We expect that dimensionality reduction will allow larger number of antecedent features to be examined and eventually provide much higher accuracies measured by detection rates/false alarm rates per class.

\section{Discussions}
\label{sec:fut}

\subsection{Industrial communication networks in the TEP benchmark}
Liu et al. proposed in \cite{liu2016simulation} a description of how a communication network could be applied in TEP.
They also proposed a  more complete analysis of Industrial IoT settings, exemplified by TEP, in \cite{liu2019wireless}.
Other recent work focusing on how 5G could be employed in TEP for fault detection and diagnosis \cite{hu20205g}.
We expect to extend those works based on the proposed 3-layer model, where we can focus on the following aspects.

\textbf{Physical layer:} It presents mostly the communication between sensors and aggregators. In order to reach a possible massive number of sensors, wireless technologies seem to be an obvious choice, in special machine-type communications (MTC) \cite{Mahmood2019}. Besides cellular massive MTC solutions \cite{LopezTWC2018}, Long Range Low Power Networks (LPWAN) (e.g. LoRa, Sigfox, NB-IoT) recently gained attention for industrial indoor applications~\cite{liu2019wireless}. See that its high link budget is suitable for the coverage of industrial environments, where we find many floors, walls, and machinery that mitigates the signal propagation~\cite{sommer2018low}. From this category, we see LoRa and LoRaWAN with great potential for industrial scenarios. Since LoRaWAN presents an open MAC protocol, it is easier to deploy aggregators (gateways) inside factories and thus have control over the whole network. Moreover, it enabled several works that evaluated LoRa’s performance~\cite{Hoeller2019Access,santana2020hybrid,rizzi2017using}. Even though we plan on a massive communication approach for mostly of sensors, there might be cases where they do not attend the reliability or latency requirements. Then, cellular ultra-reliable low-latency communications (URLLC) might come as a solution.
    
\textbf{Data layer:} The communication is based on the aggregators storing their data into a centralized unit, preceding the data fusion process. Since aggregators serve a massive number of sensors, there is a huge amount of traffic within this layer \cite{LopezTWC2018}. Thus, we plan most traffic to be enhanced Mobile BroadBand (eMBB). Similar to layer 1, we can rely on URLLC when there is reliability and latency constraints. Note that the packet size for URLLC is very limited, thus producing a small impact on the total aggregated traffic. Coexistence of MTC traffic can be mitigated by non-orthogonal solutions as in \cite{petar2018}, or through network slicing \cite{slicing2019} where orthogonal resources are  dedicated to meet each service requirement (e.g., eMBB, URLLC and mMTC). 
Considering that communication in TEP is mostly done in uplink, the customization of grant-free access based on diversity scheduling of URLLC resources as proposed in \cite{grant_free_ref1,grant_free_ref2} could be optimized in terms of reliability and users density.   
Finally, we could alternatively employ wired/optical connection where aggregators are local to the storage unit.
    
\textbf{Decision layer:} It is highly dependent on the system topology. After the data fusion stage, the aggregator must send all compressed data to a decision controller. This is a sensitive stage, where losing one packet means giving up many compressed others. Moreover, it can lead to an inaccurate detection decision. Finally, the decision results can give automatic feedback to the machinery; thus, URLLC should be predominant.

\subsection{Imperfect wireless medium}

In industrial CPS empowered by wireless connectivity, the unreliable nature of the wireless medium introduces uncertainties in the achieved TEP fault detection performance and efficiency \cite{8703476}. The complex fading conditions in indoor factory plants -- usually rich in metallic surfaces and physical obstructions which result in high network dynamics and a harsh radio propagation environment -- involving numerous IIoT components may significantly affect the accuracy of the transmitted sensor observations to the fusion center and result in transmission failures due to the high distortion levels. Although the millimeter-wave (mmWave) technology is continuously gaining momentum in industrial indoor environments for providing high data rate, low-latency and high-reliability \cite{8377337}, mmWave frequencies (up to 100 GHz) are highly susceptible to blockage, diffraction, and scattering effects. In practical industrial deployments where no line-of-sight connectivity is possible, the installation of reconfigurable intelligent surfaces, capable of adaptively shaping the impinging radio waves based on the actual channel conditions, appears as a promising solution to circumvent the unreliability of high-frequency channels.

\subsection{Communication-Computation Trade-off}
The reliability and latency concerns related to Industrial IoT in addition to the increasing density deployment precipitate the need for new communication and computation paradigms in such environments. The increasing number of sensors and the heterogeneity of the datasets being collected pose new challenges during the data fusion and analysis. For example, the large number of sensors collecting information about the industrial processes have to transmit the collected datasets to the data fusion point, which results in high communication cost and affects the energy efficiency and computational delay. There are multiple ways to approach this problem. One approach is to move the data fusion points closer to the sensors that are acquiring the measurements and to perform the analysis in the cloud.

This approach results in lower communication overhead, and reduce the amount of raw measurements being sent throughout the network, which results in higher energy efficiency. However, there are problems associated with this solution. For example, the sensors performing the data fusion are potential single points-of-failure. These fusion points also use more energy due to the computation and communication with a large number of nodes. Therefore, depending on the energy source being used for these nodes (e.g., battery), they could potentially lead to parts of the network being disconnected from the rest of the deployment. Another possible problem with this solution is related to latency concerns. Since the data analysis is happening in the cloud, the combined delay associated with the communication and computation can not be neglected. 
For this reasons, authors have been proposing the idea of moving the processing from the cloud to the edge of the network \cite{bonomi2012fog}. This approach relies on both, fusion and analysis happening on the nodes that are very close to the acquisition sensors. However, while reducing the latency effect, this approach does not address the energy hole effect (i.e., nodes closer to the centralized fusion point drain their battery faster) \cite{liu2008energy}. 

The next logical step is to rely on in-network computing, in which data processing is distributed among the nodes of the network. An example is shown in \cite{di2018network}, in which the authors place the computational nodes of a neural network on the physical sensor nodes. The placement relies on an optimized mapping procedure that minimizes either the total transmit power or the overall transmit time. Due to its flexibility, this approach allows us to eliminate the single point-of-failure problem, helps us to reduce the communication/computation latency and enables us to distribute the energy consumption across the IoT network. By taking advantage of the in-network computing paradigms we can tweak the trade-off between communication and computation. 
\subsection{Network Slices and Business Model}
Nowadays, many wireless technologies are capable to replace wired communication in industrial applications.
However, using these new technologies in real scenarios mean an increase cost in terms of CapEx and OpEx.
So, an appropriate model that fairly distributes costs over multiple virtual operators, and also optimizes physical resource planning is introduced in \cite{business_slices_ref1}.
Here, a new model of 5G isolated network slices of multitenant Mobile Backhaul (MBH) is proposed, based on a novel pay-as-you-grow model that considers the Total-Cost-of-Ownership (TCO) and the yearly generated Return-on-Investment (ROI). 
So, new business models that are coming with the wave of 5G and beyond have a big potential to boost the application of novel wireless communication technologies beyond the technical benefits.

\section{Conclusions}
\label{sec:con}
This paper shows the potential of the 3-layer approach to design anomaly detection, where data acquisition, fusion and analytics, together with the enabling communication network and computation paradigm, are jointly studied as subsequent steps.
We presented initial results based on the TEP benchmark and we plan to extend this study to other scenarios, including a micro-grid and a car factory.
All in all, we expect to demonstrate that the proposed framework is general so that it can be applied to provide ultra-reliable rare event detection in a wide range of industrial applications.
\section*{Acknowledgments}
This work is supported by CHIST-ERA (call 2017) via FIREMAN consortium, which is funded by the following national foundations: Academy of Finland (n. 326270, n. 326301), Irish Research Council, and the Spanish Government under grant PCI2019-103780. This work is partially funded by Academy of Finland 6Genesis Flagship  (n. 318927), ee-IoT (n.319009) 
and EnergyNet Research Fellowship (n.321265/n.328869).
and was supported in part by the Research Grant from Science Foundation Ireland and the European Regional Development Fund under Grant 13/RC/2077 and the Catalan Government under grant 2017-SGR-891.

\bibliographystyle{IEEEtran}
\bibliography{ref.bib}

\end{document}